\documentclass[pre, twocolumn, superscriptaddress]{revtex4}
\usepackage{amsfonts}
\usepackage{amssymb}
\usepackage{amsmath}
\usepackage{epsfig}

\renewcommand{\Vec}[1]{\boldsymbol{\rm #1}}

\DeclareMathOperator{\sgn}{sgn}

\begin{document}
\sloppy
\title{Orientational instabilities in nematics
 with weak anchoring\\ under combined action of steady flow
 and external fields}
\author{I. Sh. Nasibullayev}
\affiliation{Institute of Physics, University of Bayreuth,
D-95440, Germany} \affiliation{Institute of Mechanics, Ufa Branch,
Russian Academy of Sciences, Prosp. Oktabrya 71, 450054 Ufa,
Russia}
\author{O. S. Tarasov}
\affiliation{Institute of Molecule and Crystal Physics, Ufa Research
 Center RAS, Prosp. Oktyabrya 151, 450075 Ufa, Russia}
\author{A. P. Krekhov}
\affiliation{Institute of Physics, University of Bayreuth, D-95440, Germany}
\affiliation{Institute of Molecule and Crystal Physics, Ufa Research
 Center RAS, Prosp. Oktyabrya 151, 450075 Ufa, Russia}
\author{L. Kramer}
\affiliation{Institute of Physics, University of Bayreuth, D-95440, Germany}
\date{\today}
\begin{abstract}
We study the homogeneous and the spatially periodic instabilities
in a nematic liquid crystal layer subjected to steady plane
{\em Couette} or {\em Poiseuille} flow. The initial
director orientation is perpendicular to the flow plane.
Weak anchoring at the confining plates and the influence of the
external {\em electric} and/or {\em magnetic} field are taken into account.
Approximate expressions for the critical shear rate
are presented and compared with semi-analytical solutions in case of
Couette flow and numerical solutions of the full set of nematodynamic
equations for Poiseuille flow. In particular the dependence of the
type of instability and the threshold on the azimuthal
and the polar anchoring strength and external fields is analysed.
\end{abstract}
\maketitle
\section{Introduction}
Nematic liquid crystals (nematics) represent the simplest
anisotropic fluid. The description of the dynamic behavior of the
nematics is based on well established equations. The description
is valid for low molecular weight materials as well as nematic
polymers.

The coupling between the preferred molecular orientation (director
$\Vec{\Hat{n}}$) and the velocity field leads to interesting flow
phenomena. The orientational dynamics of nematics in flow strongly
depends on the sign of the ratio of the Leslie viscosity
coefficients $\lambda = \alpha_3 / \alpha_2$.

In typical low molecular weight nematics $\lambda$ is positive
({\em flow-aligning materials}). The case of the initial director
orientation perpendicular to the flow plane has been clarified in
classical experiments by Pieranski and Guyon
\cite{Pieranski:SSC:1973, Pieranski:PRA:1974} and theoretical
works of Dubois-Violette and Manneville (for an overview see
\cite{PF:Ch4}). An additional external magnetic field could be
applied along the initial director orientation. In Couette flow
and low magnetic field there is a homogeneous instability
\cite{Pieranski:SSC:1973}. For high magnetic field the homogeneous
instability is replaced by a spatially periodic one leading to
rolls \cite{Pieranski:PRA:1974}. In Poiseuille flow, as in Couette
flow, the homogeneous instability is replaced by a spatially
periodic one with increasing magnetic field
\cite{Manneville:JPh:1979}. All these instabilities are
stationary.

Some nematics (in particular near a nematic-smectic transition)
have negative $\lambda$ ({\em non-flow-aligning materials}). For
these materials in steady flow and in the geometry where the
initial director orientation is perpendicular to the flow plane
only spatially periodic instabilities are expected
\cite{Pieranski:CPh}. These materials demonstrate also tumbling
motion \cite{Cladis:PRL:1975} in the geometry where the initial
director orientation is perpendicular to the confined plates that
make the orientational behavior quite complicated.

Most previous theoretical investigations of the orientational
dynamics of nematics in shear flow were carried out under the
assumption of strong anchoring of the nematic molecules at the
confining plates. However, it is known that there is substantial
influence of the boundary conditions on the dynamical properties
of nematics in hydrodynamic flow \cite{Kedney:LC:V24:P613:Y1998,
Nasibullayev:MCLC:V351:P395:Y2000,
Tarasov:LC:V28:N6:P833:Y2001,Nasibullayev:CR:2001}. Indeed, the
anchoring strength strongly influences the orientational behavior
and dynamic response of nematics under external electric and
magnetic fields. This changes, for example, the switching times in
bistable nematic cells \cite{Kedney:LC:V24:P613:Y1998}, which play
an important role in applications \cite{Chigrinov:1999}. Recently
the influence of the boundary anchoring on the homogeneous
instabilities in steady flow was investigated theoretically
\cite{Tarasov:LC:V28:N6:P833:Y2001}.

In this paper we study the combined action of steady flow (Couette
and Poiseuille) and external fields (electric and magnetic) on the
orientational instabilities of the nematics with initial
orientation perpendicular to the flow plane. We focus on {\em
flow-aligning} nematics. The external electric field is applied
across the nematic layer and the external magnetic field is
applied perpendicular to the flow plane. We analyse the influence
of weak azimuthal and polar anchoring and of external fields on
both homogeneous and spatially periodic instabilities.

In section II the formulation of the problem based on the standard
set of Ericksen-Leslie hydrodynamic equations
\cite{Leslie:MCLC:1976} is presented. Boundary conditions and the
critical Fre\'edericksz field in case of weak anchoring are
discussed. In section III  equations for the homogeneous
instabilities are presented. Rigorous semi-analytical expressions
for the critical shear rate $a_c^2$ for Couette flow (section III
A), the numerical scheme for finding $a_c^2$ for Poiseuille flow
(section III B) and approximate analytical expressions for both
types of flows (section III C) are presented. In section IV the
analysis of the spatially periodic instabilities is given and in
section V we discuss the results. In particular we will be
interested in the boundaries in parameter space (anchoring
strengths, external fields) for the occurrence of the different
types of instabilities.

\section{Basic equations}

\begin{figure}
 \epsfig{file=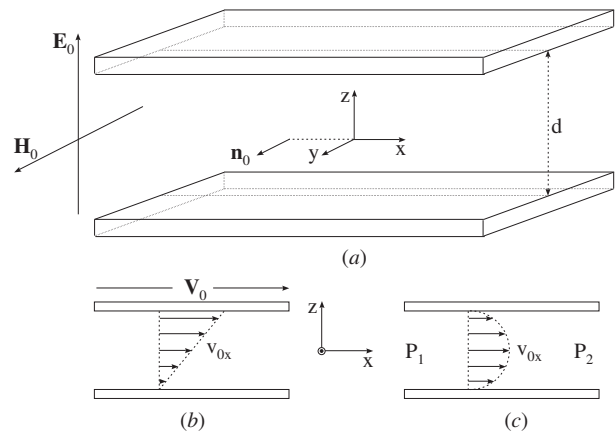,width=8cm}
 \caption{Geometry of NLC cell ($a$). Couette ($b$) and Poiseuille ($c$) flows.}
 \label{fig:cell}
\end{figure}

Consider a nematic layer of thickness $d$ sandwiched between two
infinite parallel plates that provide weak anchoring
(Fig. \ref{fig:cell} $a$). The origin of the Cartesian coordinates is
placed in the middle of the layer with the $z$ axis perpendicular
to the confining plates ($z = \pm d / 2$ for the upper/lower plate).
The flow is applied along $x$.
Steady Couette flow is induced by moving the upper plate with
a constant speed (Fig. \ref{fig:cell} $b$). Steady Poiseuille flow is induced
by applying a constant pressure difference along $x$ (Fig. \ref{fig:cell} $c$).
An external electric field $E_0$ is applied along $z$
and a magnetic field $H_0$ along $y$.

The nematodynamic equations have the following form \cite{deGennes}
\begin{eqnarray}
\label{eqn:EL:velocity}
&&\rho (\partial_t + \Vec{v} \cdot \Vec{\nabla}) v_i = - p_{,i} + [T^v_{ji} + T^e_{ji}]_{,j},\\
\label{eqn:EL:director}
&&\gamma_1 \Vec{N} = - (1 - \Vec{n} \Vec{n} \cdot) (\gamma_2 A \cdot \Vec{n} + \Vec{h}),
\end{eqnarray}
where $\rho$ is the density of the NLC and $p_{,i} = \Delta P /
\Delta x$ the pressure gradient; $\gamma_1 = \alpha_3 - \alpha_2$
and $\gamma_2 = \alpha_3 + \alpha_2$ are rotational viscosities;
$\Vec{N} = \Vec{n}_{,t} + \Vec{v} \cdot \Vec{\nabla} \Vec{n} -
(\nabla \times \Vec{v})  \times \Vec{n}/ 2$ and $A_{ij} = (v_{i,j}
+ v_{j,i}) / 2$, $h_i = \delta F / \delta n_i$. The notation
$f_{,i}\equiv \partial_i f$ is used throughout. The viscous tensor
$T^v_{ij}$ and elastic tensor $T^e_{ij}$ are
\begin{eqnarray}
&T^v_{ij} = &\alpha_1 n_i n_j A_{km} n_k n_m + \alpha_2 n_i N_j +
 \alpha_3 n_j N_i \nonumber\\
&& + \alpha_4 A_{ij} + \alpha_5 n_i n_k A_{ki} + \alpha_6 A_{ik} n_k n_j,\\
&T^e_{ij} = & - \frac{\partial F}{\partial n_{k,i}} n_{k,j},
\end{eqnarray}
where $\alpha_i$ are the Leslie viscosity coefficients. The bulk free energy density $F$ is
\begin{eqnarray}
&F = & \frac12 \Bigl\{ K_{11} (\nabla \cdot \Vec{n})^2 + K_{22} [\Vec{n} \cdot
 (\nabla \times \Vec{n})]^2 \nonumber\\
&& + K_{33} [\Vec{n} \times (\nabla \times \Vec{n})]^2 -
 \varepsilon_0 \varepsilon_a (\Vec{n} \cdot \Vec{E}_0)^2 \\
&& - \mu_0 \chi_a (\Vec{n} \cdot \Vec{H}_0)^2
 \Bigr\}.\nonumber
\end{eqnarray}
Here $K_{ii}$ are the elastic constants, $\varepsilon_a$ the anisotropy of
the dielectric permittivity and $\chi_a$ is the anisotropy of the magnetic susceptibility.

In addition one has the normalization equation
\begin{equation}
\label{eqn:normalization}
\Vec{n} = 1
\end{equation}
and incompressibility condition
\begin{equation}
\label{eqn:incompressibility}
\nabla \cdot \Vec{v} = 0.
\end{equation}
The basic state solution of equations \eqref{eqn:EL:velocity} and \eqref{eqn:EL:director} has the following form
\begin{equation}
\label{eqn:base}
\Vec{n}_0=(0,\:1,\:0),\:\Vec{v}_0=(v_{0x}(z),\:0,\:0),
\end{equation}
where $v_{0x}=V_0(1/2+z/d)$ for Couette and
$v_{0x} = (\Delta P/\Delta x)[d^2 / \alpha_4][1 / 4 - z^2 / d^2]$ for Poiseuille flow.

In order to investigate the stability of the basic state \eqref{eqn:base} with
respect to small perturbations we write:
\begin{equation}
\label{def:hom}
 \Vec{n}=\Vec{n}_0+\Vec{n}_1(z) e^{\sigma t} e^{i q y},\:
 \Vec{v}=\Vec{v}_0+\Vec{v}_1(z) e^{\sigma t} e^{i q y};
\end{equation}
We do not expect spatial variation along $x$ for steady flow. The
case $q = 0$ corresponds to a homogeneous instability. Here we
analyse stationary bifurcations, thus the threshold condition is
$\sigma = 0$.

Introducing the dimensionless quantities in terms of layer
thickness $d$ (typical length) and director relaxation time
$\tau_d = (-\alpha_2) d^2 / K_{22}$ (typical time) the linearised
equations \eqref{eqn:EL:velocity} and \eqref{eqn:EL:director} can
be rewritten in the form
\begin{subequations}
\label{eqn:rolls}
\begin{align}
\label{eqn:rolls:1}&(\eta_{13} - 1) q^2 S n_{1z} + i q (\eta_{13} q^2 - \partial_z^2) v_{1x} = 0,\\
\label{eqn:rolls:2}&\partial_z [\eta_{52} q^2 + (1 - \eta_{32}) \partial_z^2] (S n_{1x}) \nonumber\\
&\quad + (\eta_{12} q^4 - \eta_{42} q^2 \partial_z^2 + \partial_z^4) v_{1y} = 0,\\
\label{eqn:rolls:3}&(\partial_z^2 - k_{32} q^2 - h) n_{1x} + S n_{1z} + i q v_{1x} = 0,\\
\label{eqn:rolls:4}&\partial_z(k_{12} \partial_z^2 - k_{32} q^2 - h + k_{12} e) n_{1z} \nonumber\\
&\quad + \lambda \partial_z (S n_{1x}) - (q^2 + \lambda \partial_z^2) v_{1y}= 0,\\
\label{eqn:rolls:5}&v_{1z,z} = - i q v_{1y}.
\end{align}
\end{subequations}
where $\eta_{ij} = \eta_i / \eta_j$, $\eta_1 = (\alpha_4 +
\alpha_5 - \alpha_2) / 2$, $\eta_2 = (\alpha_3 + \alpha_4 +
\alpha_6) / 2$, $\eta_3 = \alpha_4 / 2$, $\eta_4 = \alpha_1 +
\eta_1 + \eta_2$, $\eta_5 = - (\alpha_2 + \alpha_5) / 2$, $k_{ij}
= K_{ii} / K_{jj}$, $\lambda = \alpha_3 / \alpha_2$, $h = \pi^2
H_0^2 / H_F^2$, $e = \sgn(\varepsilon_a) \pi^2 E_0^2/ E_F^2$ and
$H_F = (\pi / d) \sqrt{K_{22} / (\mu_0 \chi_a)}$, $E_F = (\pi / d)
\sqrt{K_{11} / (\varepsilon_0 |\varepsilon_a|)}$ are the critical
Fr\'eedericksz transition fields for strong anchoring.

For the shear rate $S$ one has, for Couette flow,
\begin{equation}
S = a^2,\: a^2 = \dfrac{V_0 \tau_d}{d}
\end{equation}
and for Poiseuille flow
\begin{equation}
S = -a^2 z,\: a^2 = -\dfrac{\Delta P}{\Delta x} \dfrac{\tau_d d}{\eta_3}.
\end{equation}

The anchoring properties are characterised by a surface energy per
unit area, $F_s$, which has a minimum when the director at the
surface is oriented along the {\em easy} axis (parallel to the $y$
axis in our case). A phenomenological expression for the surface
energy $F_s$ can be written in terms of an expansion with respect
to $(\Vec{n} - \Vec{n}_0)$. For small director deviations from the
easy axis one obtains
\begin{equation}
\label{F_s}
F_s = \frac12 W_a n_{1x}^2 + \frac12 W_p n_{1z}^2,\quad
 W_a > 0,\: W_p > 0,
\end{equation}
where $W_a$ and $W_p$ are the ``azimuthal'' and ``polar''
anchoring strengths, respectively. $W_a$ characterizes the surface
energy increase due to distortions within the surface plate and
$W_p$ relates to distortions out of the substrate plane.

The boundary conditions for the director perturbations
can be obtained from the torques balance equation
\begin{equation}
\pm \frac{\partial F}{\partial (\partial n_{1i}/\partial z)} +
 \frac{\partial F_s}{\partial n_{1i}} = 0,
\end{equation}
with ``$\pm$'' for $z = \pm d/2$.
The boundary conditions \eqref{F_s} can be rewritten in dimensionless form as:
\begin{equation}
\label{bc:director:dim}
\pm \beta_a n_{1x,z} + n_{1x} = 0,\:
\pm \beta_p n_{1z,z} + n_{1z} = 0,
\end{equation}
with ``$\pm$'' for $z = \pm 1/2$.
Here we introduced dimensionless anchoring strengths as
ratios of the characteristic anchoring length ($K_{ii} / W_i$) over the
layer thickness $d$:
\begin{equation}
\label{eqn:anchoring}
\beta_a = K_{22} / (W_a d),\: \beta_p = K_{11} / (W_p d).
\end{equation}
In the limit of strong anchoring, $(\beta_a,\:\beta_p) \to 0$, one has
$n_{1x} = n_{1z} = 0$ at $z = \pm 1/2$. For torque-free boundary
conditions, $(\beta_a,\:\beta_p)\to \infty$, one has $n_{1x,z} = n_{1z,z} = 0$ at the boundaries. From \eqref{eqn:anchoring} one can see that by changing the thickness $d$, the dimensionless anchoring strengths $\beta_a$ and $\beta_p$ can be varied with the ratio $\beta_a/\beta_p$ remaining constant.

The boundary conditions for the velocity field (no-slip) are
\begin{align}
\label{bc:vx}
&v_{1x}(z = \pm 1 / 2) = 0;\\
\label{bc:vy}
&v_{1y}(z = \pm 1 / 2) = 0;\\
\label{bc:vz}
&v_{1z}(z = \pm 1 / 2) = v_{1z,z}(z = \pm 1 / 2) = 0.
\end{align}

The existence of a nontrivial solution of the linear ordinary
differential equations \eqref{eqn:rolls} with the boundary
conditions \eqref{bc:director:dim}, (\ref{bc:vx} -- \ref{bc:vz})
gives values for the shear rate $S_0(q)$ (neutral curve). The
critical value $S_c(q_c)$, above which the basic state
\eqref{eqn:base} becomes unstable, are given by the minimum of
$S_0$ with respect to $q$.

\begin{table}
\caption{\label{table:sym:all}Symmetry properties of the solutions
of equations \eqref{eqn:rolls} under $\{z \to -z\}$.}
\begin{ruledtabular}
\begin{tabular}{ccccc}
\multicolumn{1}{c}{}&\multicolumn{2}{c}{Couette flow}&\multicolumn{2}{c}{Poiseuille flow}\\
\multicolumn{1}{c}{Perturbation}& ``odd'' & ``even'' & ``odd'' & ``even''\\
\hline
$n_{1x}$ & odd & even & odd & even\\
$n_{1z}$ & odd & even & even & odd\\
$v_{1x}$ & odd & even & odd & even\\
$v_{1y}$ & even & odd & odd & even\\
$v_{1z}$ & odd & even & even & odd\\
\end{tabular}
\end{ruledtabular}
\end{table}

The symmetry properties of the solutions of equations
\eqref{eqn:rolls} under the reflection $z \to - z$ is shown in the
Table \ref{table:sym:all}. We will always classify the solutions
by the $z$ symmetry of the $x$ component of the director
perturbation $n_{1x}$ (first row in Table I).

In case of positive $\varepsilon_a$, for some critical value of
the electric field the basic state loses its stability already in
the absence of flow ({\em Fre\'edericksz transition}). Clearly the
Fre\'edericksz transition field depends on the polar anchoring
strength. There is competition of the elastic torque $K_{11}
n_{1z,zz}$ and the field-induced torque $\varepsilon_a
\varepsilon_0 E_0^2 n_{1z}$. The solution of Eq.
\eqref{eqn:rolls:4} with $n_{1x} = 0$, $v_{1y} = 0$ for $h = 0$
has the form
\begin{equation}
\label{def:Eweak} n_{1z} = C \cos(\pi \delta z / d),
\end{equation}
where $\delta = E_F^{weak} / E_F$ and $E_F^{weak}$ is the actual
Fr\'eedericksz field.

After substituting $n_{1z}$ into the boundary conditions
\eqref{bc:director:dim} we obtain the expression for $\delta$:
\begin{equation}
\label{eqn:Eweak} \tan\dfrac{\pi \delta}2 =
 \dfrac1{\pi \beta_p \delta}.
\end{equation}

One easily sees that $\delta \to 1$ for $\beta_p \to 0$ and $\delta \to \sqrt{2 / \beta_p} / \pi$ for $\beta_p \to \infty$. For $\beta_p =1$ one gets $E_F^{weak} = 0.42 E_F$.

\section{Homogeneous instability}

In order to obtain simpler equations we use the renormalized
variables as in Ref. \cite{Tarasov:LC:V28:N6:P833:Y2001}:
\begin{align}
\label{def:renorm}
&\Tilde{S} = \beta^{-1} S,\: N_{1x} = \beta^{-1} n_{1x},\: N_{1z} = n_{1z},\: V_{1x} = \beta^{-1} v_{1x},\nonumber\\
& V_{1y} = (\beta^2 \eta_{23})^{-1} v_{1y},\: V_{1z} = (\beta^2 \eta_{23})^{-1} v_{1z}
\end{align}
with
\begin{equation}
\beta^2 = \alpha_{32} k_{21} \eta_{32},\: \alpha_{i j} = \dfrac{\alpha_i}{\alpha_j}.
\end{equation}
In the case of homogeneous perturbations ($q = 0$) Eqs.
\eqref{eqn:rolls} reduce to $V_{1z} = 0$ and
\begin{subequations}
\label{eqn:set:hom}
\begin{align}
\label{eqn:set:hom:1}
&V_{1y,zz} - (1 - \eta_{23}) (\Tilde{S} N_{1x})_{,z} = 0,\\
\label{eqn:set:hom:2}&\Tilde{S} N_{1z} - N_{1x,zz} + h N_{1x} = 0,\\
\label{eqn:set:hom:3}&\eta_{23} \Tilde{S} N_{1x} + N_{1z,zz} - V_{1y,z}  - (k_{21} h - e) N_{1z} = 0.
\end{align}
\end{subequations}

\subsection{Couette flow}

For Couette flow we can obtain the solution of \eqref{eqn:set:hom}
semi-analytically. For the ``odd'' solution one gets
\begin{align}
&N_{1x} = C_1 \sinh(\xi_1 z) + C_2 \sin(\xi_2 z),\\
&N_{1z} = C_3 \sinh(\xi_1 z) + C_4 \sin(\xi_2 z),\\
&V_{1y} = C_5 \cosh(\xi_1 z) + C_6 \cos(\xi_2 z) + C_7.
\end{align}

Taking into account the boundary conditions
(\ref{bc:director:dim}, \ref{bc:vy}) the solvability condition for
the $C_i$ (``boundary determinant'' equal to zero) gives an
expression for the critical shear rate $a_c$:
\begin{multline}
\label{eqn:couette:odd}
(h + \xi_2^2) [\xi_1 \beta_a \cosh(\xi_1 / 2) +
 \sinh(\xi_1 / 2)] \\
 \times [\xi_2 \beta_p \cos(\xi_2 / 2) + \sin(\xi_2 / 2)] \\
- (h - \xi_1^2)[\xi_2 \beta_a \cos(\xi_2 / 2) + \sin(\xi_2 / 2)] \\
 \times [\xi_1 \beta_p \cosh(\xi_1 / 2) + \sinh(\xi_1 / 2)] = 0.
\end{multline}
where
\begin{align}
&\xi_1^2 = \dfrac{[ (1 + k_{12}) h - k_{12} e] + \xi}{2 k_{12}},\\
&\xi_2^2 = \dfrac{- [(1 + k_{12}) h - k_{12} e] + \xi}{2 k_{12}},\\
&\xi = \sqrt{ [(1 - k_{12}) h - k_{12} e]^2 + 4 k_{12}^2 a^4}.
\end{align}

For the ``even'' solution one obtains:
\begin{align}
&N_{1x} = C_1 \cosh(\xi_1 z) + C_2 \cos(\xi_2 z) + C_3,\\
&N_{1z} = C_4 \cosh(\xi_1 z) + C_5 \cos(\xi_2 z) + C_6,\\
&V_{1y} = C_7 \sinh(\xi_1 z) + C_8 z.
\end{align}

The boundary conditions (\ref{bc:vx}-\ref{bc:vz}) now lead to the
following condition (``boundary determinant''):
\begin{widetext}
\begin{equation}
\label{eqn:couette:even}
\begin{vmatrix}
1 & h & \dfrac{\eta_{23}}{2}\left(\dfrac{h(h-k_{12}e)}{a^4k_{12}\eta_{23}}-1\right)\\
- \xi_2 \beta_a \sin(\xi_2 / 2) + \cos(\xi_2 / 2) &
 (h + \xi_2^2)[ - \xi_2 \beta_p \sin(\xi_2 / 2) + \cos(\xi_2 / 2)] &
 \dfrac{1 - \eta_{23}}{\xi_2}\sin(\xi_2 / 2)\\
\xi_1 \beta_a \sinh(\xi_1 / 2) + \cosh(\xi_1 / 2) &
 (h - \xi_1^2)[\xi_1 \beta_p \sinh(\xi_1 / 2) + \cosh(\xi_1 / 2)] &
 \dfrac{1 - \eta_{23}}{\xi_1}\sinh(\xi_1 / 2)
\end{vmatrix}=0.
\end{equation}
\end{widetext}

\subsection{Poiseuille flow}

In the case of Poiseuille flow the system \eqref{eqn:set:hom} with
$\Tilde{S} = - z a^2 / \beta$ admits an analytical solution only
in the absence of external fields (in terms of Airy functions)
\cite{Tarasov:LC:V28:N6:P833:Y2001}. In the presence of fields we
solve the problem numerically. In the framework of the Galerkin
method we expand $N_{1x}$, $N_{1z}$ and $V_{1y}$ in a series
\begin{align}
\label{poise:full:galerkin}
&N_{1x} = \sum\limits_{n=1}^{\infty} C_{1,n} f_n(z),\nonumber\\
&N_{1z} = \sum\limits_{n=1}^{\infty} C_{2,n} g_n(z),\\
&V_{1y} = \sum\limits_{n=1}^{\infty} C_{3,n} u_n(z),\nonumber
\end{align}
where the trial functions $f_n$, $g_n$ and $u_n$ satisfy the
boundary conditions \eqref{bc:director:dim}, \eqref{bc:vy}. For
the ``odd'' solution we write
\begin{equation}
f_n(z) = \zeta_n^o(z;\beta_a),\;
g_n(z) = \zeta_n^e(z;\beta_p),\:
u_n(z) = \nu_n^o(z)
\end{equation}
and for the ``even'' solution
\begin{equation}
f_n(z) = \zeta_n^e(z;\beta_a),\;
g_n(z) = \zeta_n^o(z;\beta_p),\:
u_n(z) = \nu_n^e(z).
\end{equation}
The functions $\zeta_n^o(z;\beta)$, $\zeta_n^e(z;\beta)$,
$\nu_n^o(z)$, $\nu_n^e(z)$ are given in Appendix A. In our
calculations we have to truncate the expansions
\eqref{poise:full:galerkin} to a finite number of modes.

After substituting \eqref{poise:full:galerkin} into the system
\eqref{eqn:set:hom} and projecting the equations on the trial functions
$f_n(z)$, $g_n(z)$ and $u_n(z)$ one gets a system of linear homogeneous algebraic equations for $\Vec{X} = \{C_{i,n}\}$ in the form
$(A - a^2 B) \Vec{X} = 0$.
We have solved this eigenvalue problem for $a^2$. The
lowest (real) eigenvalue corresponds to
the critical shear rate $a_c^2$.
According to the two types of $z$-symmetry of the solutions (and of the set
of trial functions) one obtains the threshold values of $a_c^2$ for the
``odd'' and ``even'' instability modes.
The number of Galerkin modes was chosen such that the accuracy
of the calculated eigenvalues was better than 1\% (we took ten
modes in case of ``odd'' solution and five modes for ``even'' solution).

\subsection{Approximate analytical expression for the critical shear rate}

In order to obtain an {\em easy-to-use} analytical expression for
the critical shear rate as a function of the surface anchoring
strengths and the external fields we use the lowest-mode
approximation in the framework of the Galerkin method. By
integrating \eqref{eqn:set:hom:1} over $z$ one can eliminate
$V_{1y,z}$ from \eqref{eqn:set:hom:3} which gives
\begin{equation}
\label{eqn:set:director}
\Tilde{S} N_{1x} + N_{1z,zz} + (k_{21} h - e) N_{1z} = K,
\end{equation}
where $K$ is an integration constant. Taking into account the
boundary conditions for $V_{1y}$ one has
\begin{equation}
\label{eqn:K}
K - (1 - \eta_{32}) \int\limits_{-1/2}^{1/2} S N_{1x}(z) \:dz = 0.
\end{equation}

We choose for the director components $N_{1x}$, $N_{1z}$
the one-mode approximation
\begin{equation}
\label{eqn:leading}
N_{1x} = C_1 f(z),\:
N_{1z} = C_2 g(z),
\end{equation}
Substituting \eqref{eqn:leading} into \eqref{eqn:set:hom:2} and
\eqref{eqn:set:director} and projecting the first equation on $f(z)$
and the second one on $g(z)$ we get algebraic equations for $C_i$.
The solvability condition [together with \eqref{eqn:K}]
gives the expression for the critical shear rate $a_c^2$
\begin{equation}
\label{eqn:one:ac}
a_c^2 = \sqrt{\dfrac{c_1 c_2}
 {c_3}},
\end{equation}
where $c_1 =\langle ff''\rangle  - h \langle f^2\rangle $, $c_2 =
\langle gg''\rangle  - (h/k_{12} - e) \langle g^2\rangle $, $c_3 =
\langle sfg\rangle [\langle sfg\rangle  - (1 - \eta_{23}) \langle
sf\rangle  \langle g\rangle ]$, where $\langle \dots\rangle $
denotes a spatial average
\begin{equation}
\label{def:int} \langle \dots\rangle
=\int\limits_{-1/2}^{1/2}(\dots)\:dz.
\end{equation}
The values for the integrals $\langle \dots\rangle $ are given in
Appendix B. In Table \ref{table:trial_func:appr} and Appendix A
the trial functions used are given. Equation \eqref{eqn:one:ac}
can be used for both Couette and Poiseuille flow by choosing the
function $s(z)$ [where $s(z) = 1$ for Couette flow and $s(z) = -
z$ for Poiseuille flow] and the trial functions $f(z)$ and $g(z)$
with appropriate symmetry.
\begin{table}
\caption{\label{table:trial_func:appr}Trial functions for the homogeneous solutions.}
\begin{ruledtabular}
\begin{tabular}{ccccc}
\multicolumn{1}{c}{}&\multicolumn{2}{c}{Couette flow}&\multicolumn{2}{c}{Poiseuille flow}\\
\multicolumn{1}{c}{Function}& ``odd'' & ``even'' & ``odd'' & ``even''\\
\hline
$f(z)$ & $\zeta_1^o(z;\:\beta_a)$ & $\zeta_1^e(z;\:\beta_a)$ &
  $\zeta_1^o(z;\:\beta_a)$ & $\zeta_1^e(z;\:\beta_a)$\\
$g(z)$ & $\zeta_1^o(z;\:\beta_p)$ & $\zeta_1^e(z;\:\beta_p)$ &
  $\zeta_1^e(z;\:\beta_p)$ & $\zeta_1^o(z;\:\beta_p)$\\
\end{tabular}
\end{ruledtabular}
\end{table}

For the material MBBA in the case of Couette flow the one-mode
approximation \eqref{eqn:one:ac} for the ``odd'' solution gives an
error that varies from  2.5\% to 16\% when $H_0 / H_F$ varies from
0 to 4. The ``even'' solution has an error of $0.6\% \div 8\%$ for
$0 \leqslant H_0 / H_F \leqslant 3$ and of $0.6\% \div 12\%$ for
$0 \leqslant E_0 / E_F \leqslant 0.6$.

For Poiseuille flow for ``odd'' solution the error
is $29\%$ in the absence of fields.
For the ``even'' solution the error is $12\% \div 15\%$ for
magnetic fields $0 \leqslant H_0 / H_F \leqslant 0.5$.

For both Couette and Poiseuille flow the accuracy of the formula
\eqref{eqn:one:ac} decreases with increasing field strengths.

\section{Spatially periodic instabilities}

We used for Eqs. \eqref{eqn:rolls} again the renormalized
variables \eqref{def:renorm}. The system \eqref{eqn:rolls} has no
analytical solution. Thus we solved the problem numerically in the
framework of the Galerkin method:
\begin{eqnarray}
\label{Ansatz:rolls}
N_{1x} = e^{iqy} \sum\limits_{n=1}^{\infty} C_{1,n} f_n(z),\:
N_{1z} = e^{iqy} \sum\limits_{n=1}^{\infty} C_{2,n} g_n(z),\\
V_{1x} = e^{iqy} \sum\limits_{n=1}^{\infty} C_{3,n} u_n(z),\:
V_{1z} = e^{iqy} \sum\limits_{n=1}^{\infty} C_{4,n} w_n(z).
\end{eqnarray}

After substituting \eqref{Ansatz:rolls} into the system
\eqref{eqn:rolls} and projecting on to the trial functions
$\{f_n(z),\:g_n(z),\:u_n(z),\:w_n(z)\}$ we get a system
of linear homogeneous algebraic equations for $\Vec{X} = \{C_{i,n}\}$.
This system has the form $[A(q) - a^2(q) B(q)] \Vec{X} = 0$.
We have solved the eigenvalue problem numerically to find the marginal stability curve $a(q)$. For the numerical calculations
we have chosen the trial functions shown in Table \ref{table:trial_func:rolls} and Appendix A.
\begin{table}
\caption{\label{table:trial_func:rolls}Trial functions for the spatially periodic solutions.}
\begin{ruledtabular}
\begin{tabular}{ccccc}
\multicolumn{1}{c}{}&\multicolumn{2}{c}{Couette flow}&\multicolumn{2}{c}{Poiseuille flow}\\
\multicolumn{1}{c}{Function}& ``odd'' & ``even'' & ``odd'' & ``even''\\
\hline
$f(z)$ & $\zeta_n^o(z;\:\beta_a)$ & $\zeta_n^e(z;\:\beta_a)$ &
  $\zeta_n^o(z;\:\beta_a)$ & $\zeta_n^e(z;\:\beta_a)$\\
$g(z)$ & $\zeta_n^o(z;\:\beta_p)$ & $\zeta_n^e(z;\:\beta_p)$ &
  $\zeta_n^e(z;\:\beta_p)$ & $\zeta_n^o(z;\:\beta_p)$\\
$u(z)$ & $\nu_n^o(z)$ & $\nu_n^e(z)$ &
  $\nu_n^o(z)$ & $\nu_n^e(z)$\\
$w(z)$ & $\varsigma_n^o(z)$ & $\varsigma_n^e(z)$ &
  $\varsigma_n^e(z)$ & $\varsigma_n^o(z)$\\
\end{tabular}
\end{ruledtabular}
\end{table}

In order to get an approximate expression for the threshold we
use the leading-mode approximation in the framework of the Galerkin
method. We used the same scheme described above
for the single mode and get the following formula for the
critical shear rate:
\begin{equation}
\label{rolls:one:common}
a_c^2 = \sqrt{ \eta_{23} f_x f_z / (\tilde{\alpha}_2\tilde{\alpha}_3) },
\end{equation}
with
\begin{align}
&f_x = \langle ff''\rangle  - (q^2 k_{32} + h) \langle f^2\rangle ,\\
&f_z = \langle gg''\rangle  - (q^2 k_{31} + k_{12} h - e) \langle g^2\rangle ,\\
&\tilde{\alpha}_2 =
 [ \langle fsg\rangle  - q^2 (1 - \eta_{31}) \langle fu\rangle  \langle gsu\rangle  / \gamma ],\\
&\tilde{\alpha}_3 =
 \langle fsg\rangle  + [ \alpha_{23} q^2 \langle gw\rangle  + \alpha_3 \langle gw''\rangle  ] \\
& \quad \times [ (1 - \eta_{32}) \langle w[sf]''\rangle  - \eta_{52} q^2 \langle wsf\rangle  ] / r,\\
&\gamma = q^2 \langle uu\rangle  - \eta_{31} \langle uu''\rangle ,\\
&r = \langle ww^{(4)}\rangle  - \eta_{42} q^2 \langle ww''\rangle
+ \eta_{12} q^4 \langle ww\rangle .
\end{align}
The values of the integrals $\langle \dots\rangle $ appearing in
the expression \eqref{rolls:one:common} are given in Appendix C.

In the case of strong anchoring an approximate analytical
expression for $a_c^2 = a_c^2(q_c)$ was obtained by Manneville
\cite{Manneville:JdPh:1976:285} using test functions that satisfy
free-slip boundary conditions. The formula
\eqref{rolls:one:common} is more accurate because we chose for
$v_{1z}$ Chandrasekhar functions that satisfy the boundary
conditions \eqref{bc:vz}.

For calculations we used material parameters for MBBA. The
accuracy of \eqref{rolls:one:common} is better than 1\% for
Couette flow and better than 3\% for Poiseuille flow. Note, that
Eq. \eqref{eqn:one:ac} for the homogeneous instability is more
accurate than \eqref{rolls:one:common} for $q = 0$ because
\eqref{rolls:one:common} was obtained by solving four equations
\eqref{eqn:rolls} by approximating all variables, whereas
\eqref{eqn:one:ac} was obtained by solving the reduced equations
\eqref{eqn:set:hom} by approximating only two variables.

\section{Discussion}

For the calculations we used parameters for MBBA at 25 $^\circ$C
\cite{mp:MBBA}. Calculations were made for the range of anchoring
strengths $\beta_a = 0 \div 1$ and $\beta_p = 0 \div 1$.

\subsection{Couette flow}

\begin{figure}
  \epsfig{file=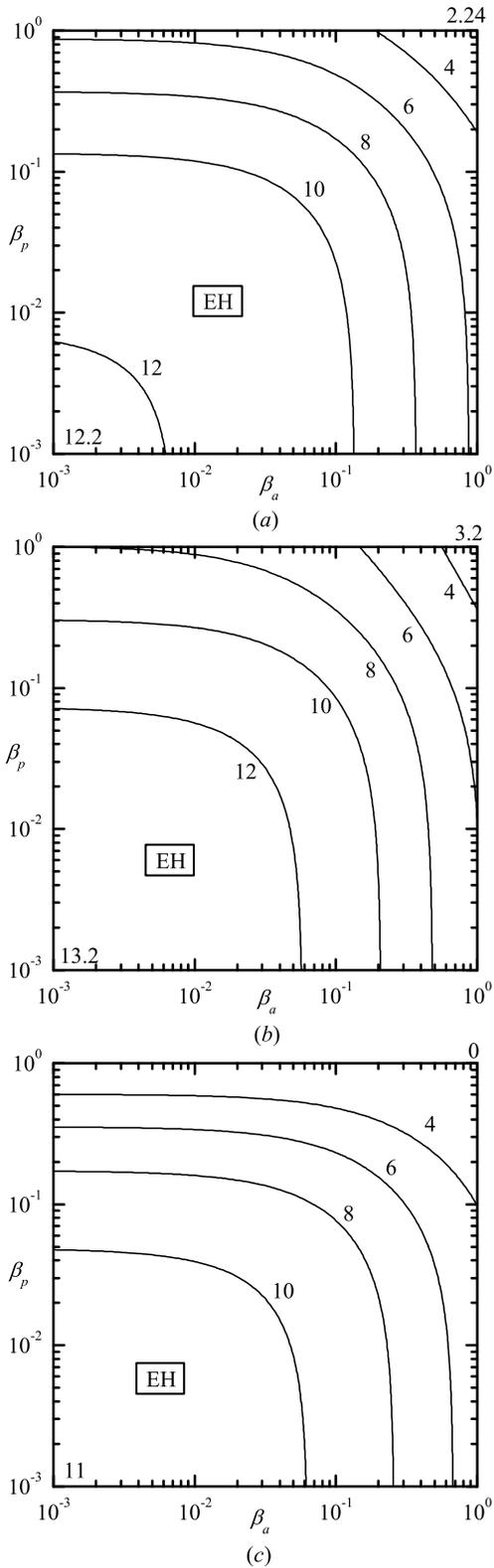,height=21cm}
  \caption{Contour plot of the critical shear rate $a_c^2$ for Couette flow
   vs. $\beta_a$ and $\beta_p$. $a$: $E_0 = 0$;
   $b$: $E_0 = E_F^{weak}$, $\varepsilon_a < 0$;
   $c$: $E_0 = E_F^{weak}$, $\varepsilon_a > 0$.
   $E_F$ is defined after Eq. \eqref{def:Eweak} and calculated in
   Eq. \eqref{eqn:Eweak}.}
  \label{fig:couette0}
\end{figure}
We found that without and with an additional electric field the
critical shear rate $a_c^2$ for the ``even'' type homogeneous
instability (EH) is systematically lower than the threshold for
other types of instability (Fig. \ref{fig:couette0}a--c). Note,
that in the presence of the field the symmetry with respect to the
exchange $\beta_a \leftrightarrow \beta_p$ is broken.

In Fig. \ref{fig:couette0} contour plots for the critical value
$a_c^2$ vs. anchoring strengths $\beta_a$ and $\beta_p$ for
different values of the electric field are shown. The differences
between $a_c^2$ obtained from the exact, semi-analytical solution
\eqref{eqn:couette:even} and from the one-mode approximation
\eqref{eqn:one:ac} are indistinguishable in the figure.

\begin{figure}
  \epsfig{file=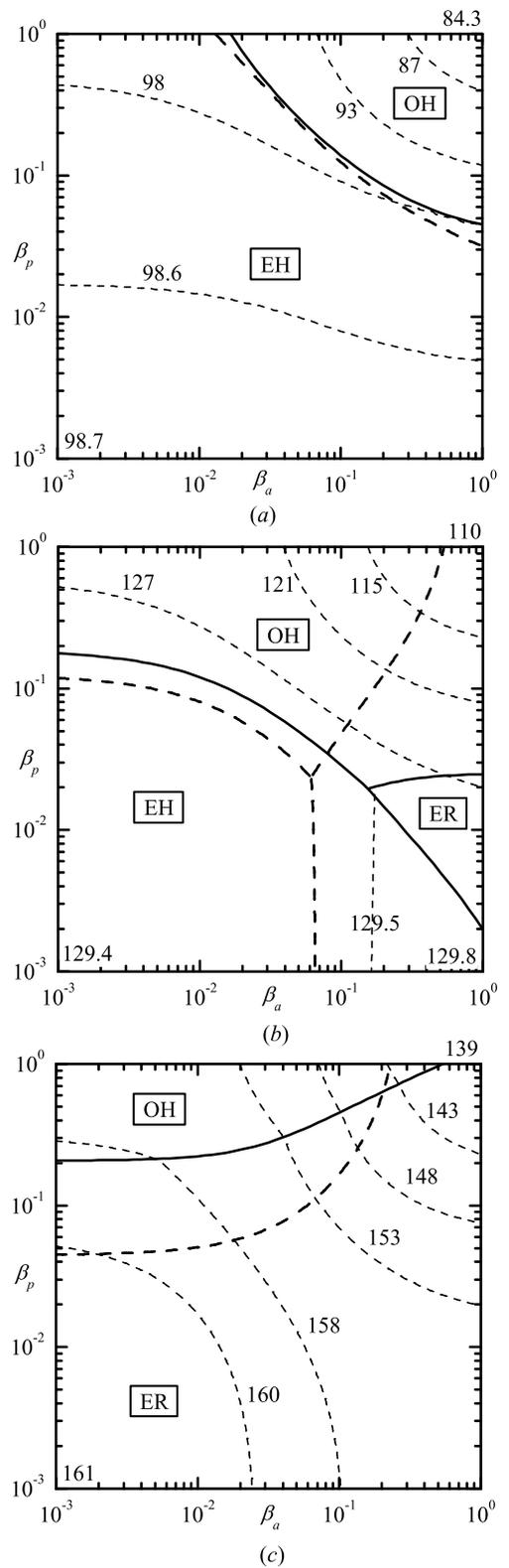,height=21cm}
  \caption{Critical shear rates and phase diagram for the instabilities under Couette flow
    with additional magnetic field.
    $a$: $H_0 / H_F = 3$; $b$: $H_0 / H_F = 3.5$; $c$: $H_0 / H_F = 4$.
    Boundaries for occurrence of instabilities are given by thick solid lines
    (full numerical) and thick dashed lines (one-mode approximation).}
  \label{fig:couette1}
\end{figure}
In Fig. \ref{fig:couette1} contour plots of $a_c^2$ (thin dashed
lines) and the boundaries where the type of instability changes
[the solid lines are obtained numerically, the thick dashed lines
from \eqref{eqn:one:ac}] for different values of magnetic field
are shown. For not too strong magnetic field in the region of weak
anchoring the ``odd'' type homogeneous instability (OH) takes
place (Fig. \ref{fig:couette1}a). Increasing the magnetic field
the OH region expands toward stronger anchoring strengths. Above
$H_0 \approx 3.2$ a region with lowest threshold corresponding to
the ``even'' roll mode (ER) appears. This region has borders with
both types of the homogeneous instability (Fig.
\ref{fig:couette1}b). With increasing magnetic field the ER region
increases (Fig. \ref{fig:couette1}c) and above $H_0 / H_F = 4$ the
ER instability has invaded the whole investigated parameter range.
For strong anchoring and $H_0 / H_F = 3.5$ the critical wave
vector is $q_c = 5.5$. It increases with increasing magnetic field
and decreases with decreasing anchoring strengths. With increasing
magnetic field the threshold for the EH instability becomes less
sensitive to the surface anchoring. Leslie has pointed out (using
an approximate analytical approach) that for strong anchoring a
transition from a homogeneous state without transverse flow (EH)
to one with such flow (OH) as the magnetic field is increased is
not possible in MBBA because of the appearance of the ER type
instability \cite{Leslie:MCLC:1976}. This is consistent with our
results. We find that the EH--OH transition in MBBA is possible
only in the region of weak anchoring (Figs.
\ref{fig:couette1}a--c).

\begin{figure}
  \epsfig{file=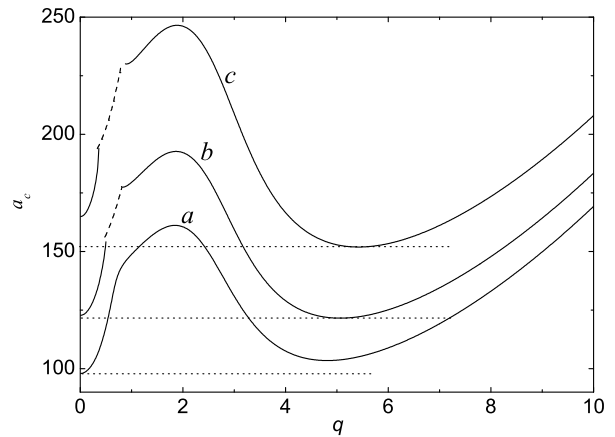,width=8cm}
  \caption{$a_c$ vs. $q$. Couette flow, $\beta_a = 0.1$, $\beta_p = 0.1$.
    $a$: $H_0 / H_F = 3$; $b$: $H_0 / H_F = 3.4$; $c$: $H_0 / H_F = 4$.}
  \label{fig:couette2}
\end{figure}
In Fig. \ref{fig:couette2} marginal stability curves for different
values of the magnetic field and fixed anchoring strengths is
shown (solid line for ER and dashed lines for OR). There are
always two minima for the even mode; one of them at $q = 0$ that
corresponds to the homogeneous instability EH. For small magnetic
field the absolute minimum is at $q = 0$ (line a). The OR curve is
systematically higher than ER. In a small range of $q$ (dotted
lines) a stationary ER solution does not exist but we have OR
instead. With increasing magnetic field the critical amplitude for
the EH minimum ($q = 0$) increases more rapidly then the one for
the ER minimum ($q \neq 0$) so that for $H_0 / H_F > 3.4$ the ER
solution is realized (lines b and c). The range of $q$ where ER is
replaced by OR expands with increasing magnetic field.

For the ER instability in the absence of fields and strong
anchoring we find $a_c^2 = 12.15$ from the semi-analytical
expression \eqref{eqn:couette:even} as well as from the one-mode
approximation \eqref{eqn:one:ac} and also \eqref{rolls:one:common}
with $q = 0$. The only available experimental value for $a_c^2$ is
$6.3 \pm 0.3$ \cite{Pieranski:SSC:1973}. We suspect that the
discrepancy is due to deviations from the strong anchoring limit
and the difference in the material parameters of the substance
used in the experiment. Assuming $\beta_a \ll 1$ one would need
$\beta_p \approx 1$ to explain the experimental value.

\subsection{Poiseuille flow}

\begin{figure}
 \epsfig{file=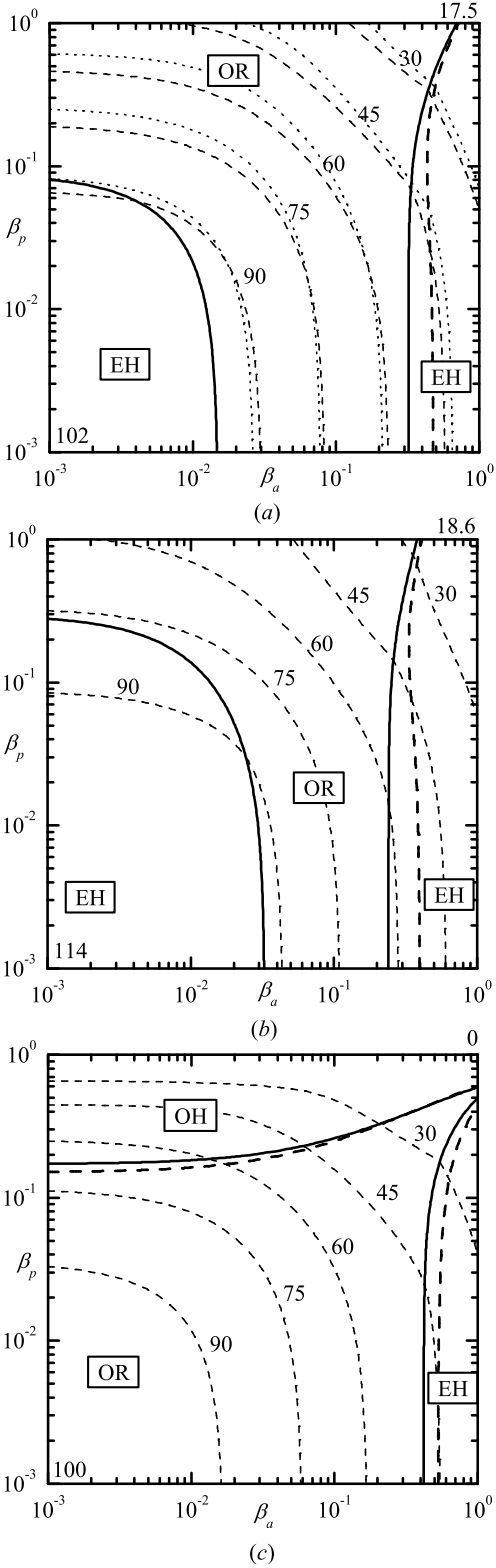,height=21cm}
 \caption{Critical shear rates and phase diagram for the instabilities in
  Poiseuille flow. $a$: $E_0 = 0$;
  $b$: $E_0 = E_0^{weak}$, $\varepsilon_a < 0$;
  $c$: $E_0 = E_0^{weak}$, $\varepsilon_a > 0$.
  Thin dashed lines: full numerical threshold;
  dotted lines: one-mode approximation for threshold.
  Boundaries for occurrence of instabilities are given by thick solid lines
  (full numerical) and thick dashed lines (one-mode approximation).}
 \label{fig:poise0}
\end{figure}

In Fig. \ref{fig:poise0} the contour plot for $a_c^2$ [thin dashed
lines from the full numerical calculation, dotted lines from the
one-mode approximations \eqref{eqn:one:ac} and
\eqref{rolls:one:common}] and the boundary for the various types
of instabilities [thick solid line: numerical; thick dashed line:
\eqref{eqn:one:ac} and \eqref{rolls:one:common}] are shown. In
Poiseuille flow  the phase diagram is already very rich in the
absence of external fields. In the region of large $\beta_a$ one
has the EH instability. For intermediate anchoring strengths rolls
of type OR occur [Fig. \ref{fig:poise0}$a$]. Note, that even in
the absence of the field there is no symmetry under exchange
$\beta_a \leftrightarrow \beta_p$, contrary to Couette flow. The
one-mode approximations \eqref{eqn:one:ac} and
\eqref{rolls:one:common} not give the transition to EH for strong
anchoring. Here we should note that in that region the difference
between the EH and the OR instability thresholds is only about
5\%. By varying material parameters [increase $\alpha_2$ by 10\%
or decrease $\alpha_3$ by 20\% or $\alpha_5$ by 25\% or $K_{33}$
by 35\%] it is possible to change the type of instability in that
region.

Application of an electric field leads for $\varepsilon_a < 0$
($\varepsilon_a > 0$) to expansion (contraction) of the EH region
[Figs. \ref{fig:poise0}$b$ and \ref{fig:poise0}$c$]. At $E_0 / E_F
= 1$ and $\varepsilon_a < 0$ rolls vanish completely and the EH
instability occurs in the whole area investigated. For
$\varepsilon_a > 0$ the instability of OH type appears in the
region of large $\beta_p$. In this case, increasing the electric
field from $E_F^{weak}$ to $E_F$ cause an expansion of the OH
region. Note that for $\beta_p > 1$, which is in the OH region,
the Fre\'edericksz transition occurs first .

\begin{figure}
 \centering
 \epsfig{file=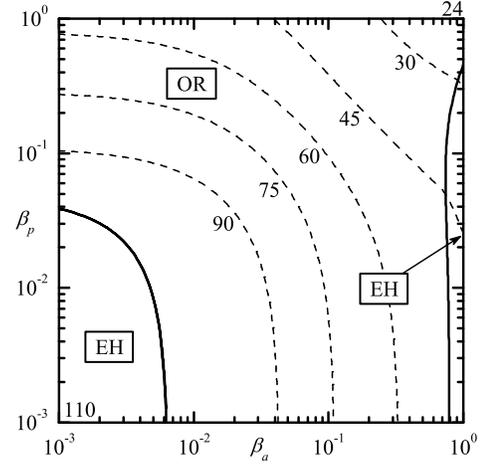,width=6.3cm}
 \caption{Phase diagram for the instabilities under Poiseuille flow
  with an additional magnetic field ($H_0 / H_F = 0.4$).}
 \label{fig:poise1}
\end{figure}
An additional magnetic field suppresses the homogeneous
instability (Fig. \ref{fig:poise1}). Above $H_0 / H_F \approx 0.5$
the OR instability (Fig. \ref{fig:poise1}) occurs for all
anchoring strengths investigated.

The wave vector $q_c$ in the absence of fields is $1.4$.
Application of an electric field decreases $q_c$ whereas the
magnetic field increases $q_c$. The wave vector decreases with
decreasing anchoring strengths.

In the absence of fields and strong anchoring we find for the EH
instability $a_c = 102$ [Eq. \eqref{eqn:one:ac} gives 110 and Eq.
\eqref{rolls:one:common} with $q = 0$ gives 130]. The experimental
value is 92 \cite{Guyon:JdPh:1975}. Thus, theoretical calculations
and experimental results are in good agreement. Note, that in the
experiments \cite{Guyon:JdPh:1975} actually not steady but
oscillatory flow with very low frequency was used ($f = 5 \cdot
10^{-3}$ Hz).

In summary, the orientational instabilities for both steady
Couette (semi-analytical for homogeneous instability and numerical
for rolls) and Poiseuille flow (numerical) were analysed
rigorously  taking into account weak anchoring and the influence
of external fields. Easy-to-use expressions for the threshold of
all possible types of instabilities were obtained and compared
with the rigorous calculations. In particular the region in
parameter space where the different types of instabilities
occurred were determined.

\acknowledgments

Financial support from DFG (project Kr690/22-1 and EGK
``Non-equilibrium phenomena and phase transition in complex
systems'').

\appendix
\section{Trial functions}

In the calculations we used the following set of trial functions:
\begin{align*}
&\zeta_n^o(z;\beta) = \sin(2 n \pi z) +
  2 n \pi \beta \sin([2 n - 1] \pi z),\\
&\zeta_n^e(z;\beta) = \cos([2 n - 1] \pi z) +
 (2 n - 1) \pi \beta \cos(2 [n - 1] \pi z),\\
&\nu_n^o(z) = \sin(2 n \pi z),\:
  \nu_n^e(z) = \cos([2 n - 1] \pi z),\\
&\varsigma_n^o(z) =
 \dfrac{\sinh(\lambda_{2 n} z)}{\sinh(\lambda_{2 n} / 2)} -
 \dfrac{\sin(\lambda_{2 n} z)}{\sin(\lambda_{2 n} / 2)},\\
&\varsigma_n^e(z) =
 \dfrac{\cosh(\lambda_{2 n - 1} z)}{\cosh(\lambda_{2 n - 1} / 2)} -
 \dfrac{\cos(\lambda_{2 n - 1} z)}{\cos(\lambda_{2 n - 1} / 2)},
\end{align*}
$\varsigma_n^o(z)$ and $\varsigma_n^e(z)$ are the Chandrasekhar
functions and $\lambda_n$ are the roots of the corresponding
characteristic equations \cite{Chandrasekhar:1993}.

\section{Integrals for the homogeneous instability}

\subsection{Couette flow}
\noindent ``Odd'' solution: $\langle sf\rangle  = \langle g\rangle
= 0$, $\langle f^2\rangle  = (3 + 32 \beta_a + 12 \pi^2 \beta_a^2)
/ 6$, $\langle g^2\rangle  = (3 + 32 \beta_p + 12 \pi^2 \beta_p^2)
/ 6$, $\langle sfg\rangle  = [3 + 16 (\beta_a + \beta_p) + 12
\pi^2 \beta_a \beta_p] / 6$, $\langle ff''\rangle  = - 2 (3 + 20
\beta_a + 3 \pi^2 \beta_a^2) / 3$, $\langle gg''\rangle  = - 2 (3
+ 20 \beta_p + 3 \pi^2 \beta_p^2) / 3$.

\noindent ``Even'' solution: $\langle sf\rangle  = (2 + \pi^2
\beta_a) / \pi$, $\langle g\rangle  = (2 + \pi^2 \beta_p) / \pi$,
$\langle f^2\rangle  = (1 + 8 \beta_a + 2 \pi^2 \beta_a^2) / 2$,
$\langle g^2\rangle  = (1 + 8 \beta_p + 2 \pi^2 \beta_p^2) / 2$,
$\langle sfg\rangle  = [1 + 4 (\beta_a + \beta_p) + 2 \pi^2
\beta_a \beta_p] / 2$, $\langle ff''\rangle  = \pi^2 (1 + 4
\beta_a) / 2$, $\langle gg''\rangle  = \pi^2 (1 + 4 \beta_p) / 2$.

\subsection{Poiseuille flow}

\noindent ``Odd'' solution: $\langle sf\rangle  = - (1 + 8
\beta_a) / (2 \pi)$, $\langle g\rangle  = - (2 + \pi^2 \beta_p) /
\pi$, $\langle f^2\rangle  = (3 + 32 \beta_a + 12 \pi^2 \beta_a^2)
/ 6$, $\langle g^2\rangle  = (1 + 8 \beta_p + 2 \pi^2 \beta_p^2) /
2$, $\langle sfg\rangle  = - [16 + 9 \pi^2 (\beta_a + \beta_p) +
72 \pi^2 \beta_a \beta_p] /
 (18 \pi^2)$,
$\langle ff''\rangle  = - 2 \pi^2 (3 + 20 \beta_a + 3 \pi^2
\beta_a^2) / 3$, $\langle gg''\rangle  = - \pi^2 (1 + 4 \beta_p) /
2$.

\noindent ``Even'' solution: $\langle sf\rangle  = I(g) = 0$,
$\langle f^2\rangle  = (1 + 8 \beta_a + 2 \pi^2 \beta_a^2) / 2$,
$\langle g^2\rangle  = (3 + 32 \beta_p + 12 \pi^2 \beta_p^2) / 6$,
$\langle sfg\rangle  = - [16 + 9 \pi^2 (\beta_a + \beta_p) + 72
\pi^2 \beta_a \beta_p] /
 (18 \pi^2)$,
$\langle ff''\rangle  = - \pi^2 (1 + 4 \beta_a) / 2$, $\langle
gg''\rangle  = - 2 \pi^2 (3 + 20 \beta_p + 3 \pi^2 \beta_p^2)$.

\section{Integrals for the spatially periodic instability}

\subsection{Couette flow}

\noindent ``Odd'' solution: $\langle wsf\rangle  \approx 0.69043 +
3.2870 \beta_a$, $\langle w[sf]''\rangle  \approx -27.258 - 32.441
\beta_a$, $\langle f^2\rangle  = (3 + 32 \beta_a + 12 \pi^2
\beta_a^2) / 6$, $\langle ff''\rangle  = - \pi^2 (6 + 40 \beta_a +
6 \pi^2 \beta_a^2) / 3$, $\langle fsg\rangle  = (3 + 16 (\beta_a +
\beta_p) + 12 \pi^2 \beta_a \beta_p) / 6$, $\langle gsu\rangle  =
(3 + 16 \beta_p) / 6$, $\langle g^2\rangle  = (3 + 32 \beta_p + 12
\pi^2 \beta_p^2) / 6$, $\langle gg''\rangle  = - \pi^2 (6 + 40
\beta_p + 6 \pi^2 \beta_p^2) / 3$, $\langle u^2\rangle  = 1 / 2$,
$\langle uu''\rangle  = - 2 \pi^2$, $\langle fu\rangle  = (3 + 16
\beta_a) / 6$, $\langle w^2\rangle  = 1$, $\langle ww''\rangle
\approx -46.050$, $\langle ww^{(4)}\rangle  = 3803.5$, $\langle
gw\rangle  \approx 0.69043 + 3.2870 \beta_p$, $\langle gw''\rangle
\approx -27.257 - 32.441 \beta_p$.

\noindent ``Even'' solution: $\langle wsf\rangle  \approx 0.69739
+ 2.6102 \beta_a$, $\langle w[sf]''\rangle  \approx -6.8828$,
$\langle f^2\rangle  = (1 + 8 \beta_a + 2 \pi^2 \beta_a^2) / 2$,
$\langle ff''\rangle  = - \pi^2 (1 + 4 \beta_a) / 2$, $\langle
fsg\rangle  = (1 + 4 (\beta_a + \beta_p) + 2 \pi^2 \beta_a
\beta_p) / 2$, $\langle gsu\rangle  = (1 + 4 \beta_p) / 2$,
$\langle g^2\rangle  = (1 + 8 \beta_p + 2 \pi^2 \beta_p^2) / 2$,
$\langle gg''\rangle  = - \pi^2 (1 + 4 \beta_p) / 2$, $\langle
u^2\rangle  = 1 / 2$, $\langle uu''\rangle  = - 2 \pi^2$, $\langle
fu\rangle  = (1 + 4 \beta_a) / 2$, $\langle w^2\rangle  = 1$,
$\langle ww''\rangle  \approx-12.303$, $\langle ww^{(4)}\rangle
\approx 500.56$, $\langle gw\rangle  \approx 0.69738 + 2.6102
\beta_p$, $\langle gw''\rangle  \approx -6.8828$.

\subsection{Poiseuille flow}

\noindent ``Odd'' solution: $\langle wsf\rangle  \approx -0.10292
- 0.49816 \beta_a$, $\langle w[sf]''\rangle  \approx -0.87673 -
22.615 \beta_a$, $\langle f^2\rangle  = (3 + 32 \beta_a + 12 \pi^2
\beta_a^2) / 6$, $\langle ff''\rangle  = - \pi^2 (6 + 40 \beta_a +
6 \pi^2 \beta_a^2) / 3$, $\langle fsg\rangle  =
  - (16 + 9 \pi^2 (\beta_a + \beta_p) + 72 \pi^2 \beta_a \beta_p)/(18 \pi^2)$,
$\langle gsu\rangle  = - (16 + 9 \pi^2 \beta_p) / (18 \pi^2)$,
$\langle g^2\rangle  = (1 + 8 \beta_p + 2 \pi^2 \beta_p^2) / 2$,
$\langle gg''\rangle  = - \pi^2 (1 + 4 \beta_p) / 2$, $\langle
u^2\rangle  = 1 / 2$, $\langle uu''\rangle  = - 2 \pi^2$, $\langle
fu\rangle  = (3 + 16 \beta_a) / 6$, $\langle w^2\rangle  = 1$,
$\langle ww''\rangle  \approx -12.303$, $\langle ww^{(4)}\rangle
\approx 500.56$, $\langle gw\rangle  \approx 0.69738 + 2.6102
\beta_p$, $\langle gw''\rangle  \approx -6.8828$.

\noindent ``Even'' solution: $\langle wsf\rangle  \approx -
0.12206 - 0.59694 \beta_a$, $\langle w[sf]''\rangle  \approx
4.4917$, $\langle f^2\rangle  = (1 + 8 \beta_a + 2 \pi^2
\beta_a^2)$, $\langle ff''\rangle  = - \pi^2 (1 + 4 \beta_a) / 2$,
$\langle fsg\rangle  =
 - (16 + 9 \pi^2 (\beta_a + \beta_p) + 72 \pi^2 \beta_a \beta_p)/(18 \pi^2)$,
$\langle gsu\rangle  = - (16 + 9 \pi^2 \beta_p)/(18 \pi^2)$,
$\langle g^2\rangle  = (3 + 32 \beta_p + 12 \pi^2 \beta_p^2) / 6$,
$\langle gg''\rangle  = - 2 \pi^2 (3 + 20 \beta_p + 3 \pi^2
\beta_p^2) / 3$, $\langle u^2\rangle  = 1 / 2$, $\langle
uu''\rangle  = - \pi^2 / 2$, $\langle fu\rangle  = (1 + 4 \beta_a)
/ 2$, $\langle w^2\rangle  = 1$, $\langle ww''\rangle  \approx
-46.050$, $\langle ww^{(4)}\rangle  \approx 3803.5$, $\langle
gw\rangle \approx 0.69043 + 3.2870 \beta_p$, $\langle gw''\rangle
\approx -27.257 - 32.441 \beta_p$.

\bibliography{stable,mp,anchoring}

\end{document}